\documentclass[12pt,preprint]{aastex}

\shorttitle{Non-Equilibrium $\beta$ Processes in Neutron Stars}
\shortauthors{Flores-Tuli\'an \& Reisenegger}
\def\dmu{\delta\mu}
\newcommand{\rmn}[1] {{\rm #1}}

\begin{document}
\title{Non-Equilibrium Beta Processes in Neutron Stars:
A Relationship between the Net Reaction Rate and the Total
Emissivity of Neutrinos}
\author{Sergio Flores-Tuli\'an\altaffilmark{1} and  Andreas Reisenegger\altaffilmark{2}}
\affil{Departamento de Astronom\'\i a y Astrof\'\i sica, Pontificia Universidad Cat\'olica de Chile,\\
Casilla 306, Santiago 22, Chile.}
\altaffiltext{1}{E-mail: sflores@astro.puc.cl}
\altaffiltext{2}{E-mail: areisene@astro.puc.cl}
\begin{abstract}
Several different processes could be changing the density in the
core of a neutron star, leading to a departure from $\beta$
equilibrium, quantified by the chemical potential difference
$\delta\mu\equiv\mu_n-\mu_p-\mu_e$. The evolution of this quantity
is coupled to that of the star's interior temperature $T$ by two
functions that quantify the rate at which neutrino-emitting
reactions proceed: the net reaction rate (difference between $\beta$
decay and capture rates), $\Gamma_{\rm net}(T,\delta\mu)$, and the
total emissivity (total energy emission rate in the form of
neutrinos and antineutrinos), $\epsilon_{\rm tot}(T,\delta\mu)$.
Here, we present a simple and general relationship between these
variables, ${\partial\epsilon_{\rm
tot}/\partial\delta\mu=3\Gamma_{\rm net}}$, and show that it holds
even in the case of superfluid nucleons. This relation may simplify
the numerical calculation of these quantities, including superfluid
reduction factors.
\end{abstract}


\section{INTRODUCTION}
The simplest weak interaction process that could proceed in the core
of a neutron star is the so-called direct Urca (or {\it Durca})
process, which consists of the two successive reactions, $\beta$
decay and capture.
\begin{equation}
\label{beta}n\rightarrow p+e+\bar{\nu}_e\qquad\qquad p+e\rightarrow n+\nu_e.
\end{equation}
It is the most powerful of the neutrino processes potentially
leading to the cooling of the neutron star. However, these reactions
are allowed only if each of the Fermi momenta of neutrons ($n$),
protons ($p$), and electrons ($e$) is smaller than the sum of the
two others (triangle condition). This implies a (highly uncertain)
threshold in the matter density  for the direct Urca processes.
Another reaction, which overcomes this restriction, is the so-called
modified Urca (or {\it Murca}) process, which involves an additional
spectator nucleon:
\begin{equation}
\label{MURCA}
n+N\rightarrow p+e+\bar{\nu}_e+N\qquad\qquad p+e+N\rightarrow n+\nu_e+N,
\end{equation}
where the additional nucleon $N$ can be either a neutron or a
proton.
The matter is transparent to neutrinos, which escape freely,
transporting their energy away from the star.

Furthermore, these two reactions bring nucleons into the state
of beta (or chemical) equilibrium, which determines the
concentration of neutrons and protons.
The beta equilibrium condition $\mu_n=\mu_p+\mu_e$ involving the
chemical potentials of the constituent particles implies the
equality of $\bar{\nu}_e$ and $\nu_e$ emission rates (or the net
reaction rate set equal to zero). If a slight departure from
equilibrium $\delta\mu=\mu_n-\mu_p-\mu_e$ is produced by any
external or macroscopic phenomenon that changes the density of a
fluid element, one of the two reactions in equations (\ref{beta}) or
(\ref{MURCA}) becomes more intense and changes the fraction of protons
 and neutrons towards the new equilibrium values (Le Ch\^atelier's principle).
 In the subsequent evolution, the chemical imbalance affects the stellar
interior temperature $T$ by increasing the phase space available
 to the products of the neutrino-emitting
reactions and by converting chemical energy into thermal energy, and
the temperature affects the chemical imbalance by also determining
the rate at which reactions proceed. Several authors have
investigated external processes that induce non-equilibrium $\beta$
reactions, such as radial pulsation
\citep{finzi65,finzi_wolf}, gravitational collapse
\citep{haensel92,Gourgou93}, a changing rotation rate
\citep{andreas95,andreas97,fernand05,andreas06}, and a hypothetical
time-variation of the gravitational constant \citep{jofre06}.
However, none of these has considered an adequate model for the
effects of the likely Cooper pairing (superfluidity) of nucleons
(see \citealt{andreas97} for a very rough estimate of their likely
importance).

Two basic rates are relevant in order to follow the coupled
evolution of $T$ and $\delta\mu$ (see, e.~g.,
\citealt{andreas95,fernand05}): the total (neutrino and
antineutrino) emissivity (energy per unit volume per unit time),
$\epsilon_{\rmn{tot}}=\epsilon_{\bar{\nu}}+\epsilon_\nu$, and the
net reaction rate (number of reactions or emitted lepton number per
unit volume per unit time),
$\Gamma_{\rmn{net}}=\Gamma_{\bar{\nu}}-\Gamma_\nu$. Recently,
\citet{villain05} analyzed the effect of nucleon superfluidity on
the net reaction rates, $\Gamma_{\rm net}$, calculating superfluid
reduction factors for the direct and modified Urca processes by
means of sophisticated numerical methods. Nevertheless, they did not
evaluate the total emissivity of neutrinos, $\epsilon_{\rm tot}$,
which is also required in order to compute the eventual
time-evolution of $T$ and $\delta\mu$.

In this work, we present a curious relationship between
$\epsilon_{\rmn{tot}}$ and $\Gamma_{\rmn{net}}$, both considered as
functions of the chemical imbalance parameter $\dmu$, that holds
even in the superfluid case:
\begin{equation}
\label{3sig}\frac{\partial\epsilon_{\rmn{tot}}}{\partial\dmu}=3\Gamma_{\rmn{net}}.
\end{equation}
The advantage of this relation is that, having computed one of the
functions numerically (as done by \citealt{villain05}), the other is
obtained very easily, saving time in the calculation.

In the non-superfluid case, \citet{andreas95} calculated
$\epsilon_{\rmn{tot}}$ and $\Gamma_{\rmn{net}}$ analytically. For
Durca processes,
\begin{equation}
\label{R1}
\epsilon_{\rmn{tot}}^{\rmn{Durca}}(T,\dmu)=\epsilon_0^D\left(1+\frac{1071u^2+315u^4+21u^6}{457}\right),
\end{equation}
and
\begin{equation}
\label{R2}
\Gamma_{\rmn{net}}^{\rmn{Durca}}(T,\dmu)\dmu=\epsilon_0^D\frac{714u^2+420u^4+42u^6}{457},
\end{equation}
where $u\equiv\dmu/(\pi T)$ is a dimensionless parameter, and
$\epsilon_0^D\equiv\epsilon_{\rmn{tot}}^{\rmn{Durca}}(T,0)\propto
T^6$ is the equilibrium emissivity. For Murca processes,
\begin{equation}
\label{R3}
\epsilon_{\rmn{tot}}^{\rmn{Murca}}(T,\dmu)=\epsilon_0^M\left(1+\frac{22020u^2+5670u^4+420u^6+9u^8}{11513}\right),
\end{equation}
and
\begin{equation} \label{R4}
\Gamma_{\rmn{net}}^{\rmn{Murca}}(T,\dmu)\dmu=\epsilon_0^M\frac{14680u^2+7560u^4+840u^6+24u^8}{11513},
\end{equation}
with
$\epsilon_0^M\equiv\epsilon_{\rmn{tot}}^{\rmn{Murca}}(T,0)\propto
T^8$ (see \citealt{haensel92} for precise estimates). In these two
pairs of polynomial expressions, it is easy to verify our proposed
relation (eq.~\ref{3sig}). The purpose of this paper is to show that
it is valid beyond these simple cases, encompassing, for example,
the cases of superfluid neutrons and/or protons. A brief and clear
discussion of superfluidity in neutron star was given by
\citet{yako2001b}, and a rigorous and deep analysis can be found in
\citet{lombardo}.
\section{DERIVATION}
\label{sec:Deriv} We now show that equation (\ref{3sig}) holds regardless
 of the nucleons being superfluid or not.
Along this paper we will use natural units, with $\hbar=k_B=c=1$.\\

We express the phase space factors as
\begin{equation}
d{\bf p_i}=p_i^2{\rmn{d}}p_i\,{\rmn{d}}\Omega_i= D(E_i)  {\rmn{d}}E_i\,{\rmn{d}}\Omega_i,
\end{equation}
where $E_i(p_i)$ is the particle energy,\footnote{The dispersion
relation for the superfluid nucleons can be written as
$E(p)=\mu+{\rmn{sgn}}(p\!-\!p_{F})\sqrt{v_{F}^2(p-p_{F})^2+\Delta^2}$,
where $p_{F}$ denotes the Fermi momenta. The Cooper pairing
occurring around the Fermi surfaces, allows us to approximate
$(1/2m)(p^2-p_{F}^2)\simeq v_F^2(p-p_F)^2$, where $v_F=p_F/m^*$ and
 $m\!^*$ are the Fermi velocity and effective particle mass for
superfluid nucleons, respectively.}
and
\begin{equation}
D(E_i)=p_i^2\frac{d p_i}{d E_i},
\end{equation}
is the density of states  and ${\rmn{d}}\Omega_i$ is the solid angle
element in the direction of ${\bf p_i}$.
In order to show the generality of our derivation, we left the electron
 and
nucleon densities of states\footnote{The density of states for
superfluid nucleons is $D(E)={p_F\,m^*\vert E-\mu\vert\over
\sqrt{(E-\mu)^2-\Delta^2}} \,\Theta(\vert E-\mu\vert-\Delta)$.}
expressed implicitly in the following calculation.
 For neutrinos, we will need the explicit expression
\begin{equation}\label{pnu0}
d{\bf p_\nu}=E_\nu^2\,{\rmn{d}}E_\nu d\Omega_\nu,
\end{equation}
and we may assume neutrino isotropy, so
\begin{equation}\label{pnu}
d{\bf p_\nu}=4\pi E_\nu^2\,{\rmn{d}}E_\nu.
\end{equation}

 In the interior of the neutron star, the temperature is much smaller
 than the Fermi temperature. On the other hand, the neutrino momentum
is proportional to the temperature of the star, and the other momenta
 are essentially their respective Fermi momenta (degenerate matter).
Therefore, we assume that
$\vert{\bf p_\nu}\vert \ll\vert{\bf p_i}\vert$ ($i=n,\, p,\, e$),
and neglect ${\bf p_\nu}$ in the Dirac delta function of momentum.
 This approach will allow us to integrate more easily over
the orientation of the neutrino momentum.

Using the following dimensionless  variables:
\begin{equation}\label{adim}
x_i\equiv\frac{E_i-\mu_i}{T} \quad (i=n,\,p,\,e),\quad
x_\nu\equiv\frac{E_\nu}{T}\quad \textrm{and} \quad u\equiv\frac{\delta\mu}{T},
\end{equation}
the total emissivity $\epsilon_{tot}$ can be expressed as
\begin{equation}\label{etot}
\epsilon_{\rm{tot}}(T,\delta\mu)=\frac{T^6}{(2\pi)^8}\langle|M|^2\rangle\,\widehat{\Omega}
\,\widehat{I}_-
\end{equation}
(\citealt{yako2001}, eq. 115), where $\langle|M|^2\rangle$ is the
squared transition amplitude for the $\beta$ decay and capture
processes, averaged over initial and summed over final spin states;
 furthermore averaged over the direction of the neutrino momentum, this results in
 a roughly constant value, which can be taken out of the integrals
 (for more details see \citealt{shapiro} or \citealt{yako2001}):
\begin{equation}\label{promMif}
\langle|M|^2\rangle\simeq 2G_F\cos^2\!\theta_C(1+3g_A^2),
\end{equation}
where $G_F$, $\theta_C$, and $g_A$ are the Fermi weak interaction constant, the
Cabibbo angle ($\sin\theta_C=0.231$), and the Gamow-Teller axial vector
coupling constant, respectively.
The operator $\widehat{\Omega}$ contains the integrals over the orientations of
the particle momenta
\begin{equation}\label{omega}
\widehat{\Omega}\equiv 4\pi\int\!\!\!\int\!\!\!\int
{\rmn{d}}\Omega_n{\rmn{d}}\Omega_p{\rmn{d}}\Omega_e\,\delta({\bf
p_n}\!-\!{\bf p_p}\!-\!{\bf p_e}),
\end{equation}
and $\widehat{I}_-$ integral includes the integrations over dimensionless particle energies:
\[
\widehat{I}_-\!\equiv\int_{-\infty}^\infty\mskip-10mu{\rmn{d}}x_e D_e f_e\mskip-10mu\int_{-\infty}^\infty\mskip-10mu{\rmn{d}}x_n D_n f_n \mskip-8mu\int_{-\infty}^\infty\mskip-10mu{\rmn{d}}x_p D_p f_p\int_0^\infty\!\!\!{\rmn{d}}x_\nu x_\nu^3
\]
\begin{equation}\label{I-}
\times\left[\delta(x_n\!+\!x_p\!+\!x_e\!-\!x_\nu\!+u)-\delta(x_n\!+\!x_p\!+\!x_e\!-\!x_\nu\!-\!u)\right],
\end{equation}
where the $f_i$'s (for $i=p,\, n,\, e$) are the Fermi-Dirac
distributions, $\label{FD} f_i=(e^x_i+1)^{-1}$, for some of whose
arguments the signs have been redefined when using the symmetry
property $f_i(x)=1-f_i(-x)$. The functions $D(x_i)$ are
approximately symmetric around $x=0$ for all relevant cases.

When considering situations with neutron or proton superfluidity,
the possible anisotropy of gaps (appearing in the density of states
function) in the expression  $\widehat{I}_-$ should be taken into
account in the preceding $\widehat{\Omega}$ integration.

Similarly, the net reaction rate is
\begin{equation}\label{gammanet}
\Gamma_{\rm{net}}(T,\delta\mu)=\frac{T^5}{(2\pi)^8}\langle|M|^2\rangle\,\widehat{\Omega} \,\widehat{I}_+,
\end{equation}
where
\[
\widehat{I}_+\!\equiv\int_{-\infty}^\infty\mskip-10mu{\rmn{d}}x_e D(x_e)f_e\mskip-10mu\int_{-\infty}^\infty\mskip-10mu{\rmn{d}}x_n D(x_n) f_n\mskip-8mu\int_{-\infty}^\infty\mskip-10mu{\rmn{d}}x_p D(x_p) f_p\int_0^\infty{\rmn{d}}x_\nu x_\nu^2
\]
\begin{equation}\label{I+}
\times\left[\delta(x_n\!+\!x_p\!+\!x_e\!-\!x_\nu\!+u)+\delta(x_n\!+\!x_p\!+\!x_e\!-\!x_\nu\!-\!u)\right].
\end{equation}
Note that the integral now contains only two powers of $x_\nu$.

The total emissivity (given by eq. [\ref{etot}]) depends on the
chemical imbalance $\delta\mu$ only through the dimensionless
non-equilibrium parameter $u$ contained in $\widehat{I}_-$, in the
argument of the Dirac delta function. In order to calculate the
derivative $\partial\epsilon_{\mathrm tot}/\partial\delta\mu$, we
define
\begin{equation}\label{z+-}
z_\pm\equiv x_n\!+\!x_p\!+\!x_e\!-\!x_\nu\!\pm u,
\end{equation}
so we can rewrite the derivative
\begin{equation}\label{chain}
\frac{\partial}{\partial u}\left[\delta(z_+)\!-\!\delta(z_-)\right]=
\frac{\partial}{\partial x_\nu}\left[-\delta(z_+)\!-\!\delta(z_-)\right]
\end{equation}
and do an integral by parts in order to obtain the derivative of
equation (\ref{I-})\footnote{The boundary term is of the form
$x^3/(1+e^x)\vert_0^\infty$ and vanishes in both limits.}:
\begin{equation}\label{dI-/du}
\frac{\partial\widehat{I}_-}{\partial u}\!=
\int_{-\infty}^\infty{\rmn{d}}x_e D(x_e)
f_e\int_{-\infty}^\infty{\rmn{d}}x_n D(x_n)
f_n\int_{-\infty}^\infty{\rmn{d}}x_p D(x_p)
f_p\int_0^\infty\!\!\!{\rmn{d}}x_\nu (3 x_\nu^2)
\left[\delta(z_+)+\delta(z_-)\right]=3\widehat{I}_+.
\end{equation}
Thus, in a straightforward way, we reach the proposed relation
(eq.~\ref{3sig}):
\[\frac{\partial\epsilon_{\rmn{tot}}}{\partial\dmu}=3\Gamma_{\rmn{net}}.\]

It is straightforward to verify that an analogous derivation can be made
for modified Urca processes.
\section{COOLING VS. HEATING}
\label{sec:CvsH} As a useful application of the proposed relation,
we analyze the balance between heating and cooling due to $\beta$
processes.

The net local heating rate can be written as
\begin{equation}\label{hnet}
h_{\rm{net}}=\Gamma_{\rm{net}}\delta \mu-\epsilon_{\rm{tot}},
\end{equation}
where the first term is the rate of dissipation of chemical energy,
and the second is the energy loss rate due to neutrino emission.
Using our relation (eq.~\ref{3sig}), which links
$\epsilon_{\rm{tot}}$ and $\Gamma_{\rm{net}}$, we obtain
\begin{equation}\label{hnet-compact}
h_{\rm{net}}=\frac{\delta\mu^4}{3}\frac{\partial}{\partial \delta\mu}\left(\frac{\epsilon_{\rm{tot}}}{\delta\mu^3}\right).
\end{equation}
If $\epsilon_{\rm{tot}}$ increases faster than $\delta\mu^3$, the
net heating is positive ($h_{\rm{net}}>0$), and viceversa.

The net heating rate can also be written as
\begin{equation}\label{hnet-ln}
h_{\rm{net}}=\frac{\epsilon_{\rm{tot}}}{3}\left(\frac{\partial\ln\epsilon_{\rm{tot}}}{\partial\ln\delta\mu}-3\right).
\end{equation}
Without Cooper pairing, and in the limiting case of $\delta\mu\gg
T$, we know that $\epsilon_{\rm{tot}}\propto\delta\mu^6$ for Durca
processes, and $\epsilon_{\rm{tot}}\propto\delta\mu^8$ for Murca
processes. For these cases, we easily reobtain that the total energy
released, $\Gamma_{\rm{net}}\delta\mu$, is distributed in fixed
fractions among internal heating, $h_{\rm{net}}$, and neutrino
emission, as
$h_{\rm{net}}=\epsilon_{\rm{tot}}=\frac{1}{2}\Gamma_{\rm{net}}\delta\mu$
for Durca, and $h_{\rm{net}}=\frac{5}{8}\Gamma_{\rm{net}}\delta\mu$,
$\epsilon_{\rm{tot}}=\frac{3}{8}\Gamma_{\rm{net}}\delta\mu$ for
Murca processes \citep{fernand05}.
\section{CONCLUSIONS}
\label{sec:concl} We have proven a simple, general relationship
(eq.~\ref{3sig}) between the main rates characterizing
non-equilibrium $\beta$ processes in both superfluid and
non-superfluid neutron star matter. This relation could simplify the
evaluation of these quantities in superfluid neutron star models,
complementing numerical calculations such as those of
\citet{villain05}.

We thank Francisco Claro for very helpful suggestions. We are also
grateful to Olivier Espinosa, Claudio Dib, Miguel Kiwi, Paula
Jofr\'e, and Elena Kantor for discussions that benefited the present
paper. S.~F.-T. was supported by a CONICYT PhD fellowship and A.~R.
by FONDECYT grants 1020840 and 1060644.

\end{document}